\documentstyle[12pt,epsfig,amssymb]{article}
\newcommand{\be}{\begin{equation}}
\newcommand{\ee}{\end{equation}}
\newcommand{\bea}{\begin{eqnarray}}
\newcommand{\eea}{\end{eqnarray}}

\newcommand{\dd}{\displaystyle}
\begin{document}
\def\slash#1{\setbox0=\hbox{$#1$}#1\hskip-\wd0\dimen0=5pt\advance
       \dimen0 by-\ht0\advance\dimen0 by\dp0\lower0.5\dimen0\hbox
         to\wd0{\hss\sl/\/\hss}}\def\ink {\int~{d^4k\over (2\pi)^4}~}
\def\gamh{\Gamma_H}
\def\esp #1{e^{\displaystyle{#1}}}
\def\de{\partial}
\def\eb{E_{\rm beam}}
\def\deb{\Delta E_{\rm beam}}
\def\sigm{\sigma_M}
\def\sigmmax{\sigma_M^{\rm max}}
\def\sigmmin{\sigma_M^{\rm min}}
\def\sige{\sigma_E}
\def\dsigm{\Delta\sigma_M}
\def\mh{M_H}
\def\lyear{L_{\rm year}}
\def\wstar{W^\star}
\def\zstar{Z^\star}
\def\ie{{\it i.e.}}
\def\etal{{\it et al.}}
\def\eg{{\it e.g.}}
\def\pzero{P^0}
\def\mt{m_t}
\def\mpzero{M_{\pzero}}
\def\mev{~{\rm MeV}}
\def\gev{~{\rm GeV}}
\def\gam{\gamma}
\def\lsim{\mathrel{\raise.3ex\hbox{$<$\kern-.75em\lower1ex\hbox{$\sim$}}}}
\def\gsim{\mathrel{\raise.3ex\hbox{$>$\kern-.75em\lower1ex\hbox{$\sim$}}}}
\def\ntc{N_{TC}}
\def\epem{e^+e^-}
\def\tauptaum{\tau^+\tau^-}
\def\lplm{\ell^+\ell^-}
\def\anti{\overline}
\def\mw{M_W}
\def\mz{M_Z}
\def\fbi{~{\rm fb}^{-1}}
\def\mupmum{\mu^+\mu^-}
\def\rts{\sqrt s}
\def\sigrts{\sigma_{\tiny\rts}^{}}
\def\sigrtssq{\sigma_{\tiny\rts}^2}
\def\sigrtsprime{\sigma_{E}}
\def\nsigrts{n_{\sigrts}}
\def\gampzero{\Gamma_{\pzero}}
\def\pzerop{P^{0\,\prime}}
\def\mpzerop{M_{\pzerop}}
\font\fortssbx=cmssbx10 scaled \magstep2
%\hbox to \hsize{
%
%\special{psfile=uwlogo.ps
% hscale=8000 vscale=8000
% hoffset=-12 voffset=-2}
%\hskip.5in \raise.1in
%
%$\vcenter{
%$\hbox{\fortssbx University of Florence}
%$\hbox{\fortssbx University of Geneva}
%}
%
\hfill$\vcenter{
 \hbox{\bf BARI-TH 420/01}
\hbox{\bf UGVA-DPT-2001-06-1095} }$
\begin{center}
{\Large\bf\boldmath {Effective gluon interactions in the
}}\vskip0.3cm {\Large\bf\boldmath{ Colour Superconductive Phase of
two flavor}} \vskip0.3cm {\Large\bf\boldmath{QCD}}
\\ \rm \vskip1pc {\large
R. Casalbuoni$^{a,b}$,  R. Gatto$^c$, M. Mannarelli$^{c,d,e}$
and\\ G. Nardulli$^{d,e}$}\\ \vspace{5mm} {\it{$^a$Dipartimento di
Fisica, Universit\`a di Firenze, I-50125 Firenze, Italia
\\
$^b$I.N.F.N., Sezione di Firenze, I-50125 Firenze, Italia\\
$^c$D\'epart. de Physique Th\'eorique, Universit\'e de Gen\`eve,
CH-1211 Gen\`eve 4, Suisse\\ $\dd ^d$Dipartimento di Fisica,
Universit\`a di Bari, I-70124 Bari, Italia  \\$^e$I.N.F.N.,
Sezione di Bari, I-70124 Bari, Italia }}
\end{center}
%%% ----------------------------------------------------------------------
\begin{abstract}

The gluon self-energies and dispersion laws in the color
superconducting phase of QCD with two massless flavors are
calculated using the effective theory near the Fermi surface.
These quantities are calculated at zero temperature for all the
eight gluons, those of the remaining $\dd  SU(2)$ color group and
those corresponding to the broken generators. The construction of
the effective interaction is completed with the one loop
calculation of the three- and four-point gluon interactions.
\end{abstract}
\vskip1.cm

\section{Introduction}
Color superconductivity in QCD at large densities is a rather
ancient idea \cite{others} that has recently received a new
attention in a series of papers \cite{alford}, \cite{alford1} (for
recent review see \cite{rassegne}). Both three-flavor (Color
Flavor Locking model=CFL) and two-flavor cases (2SC model) have
been studied. In this letter we consider the 2SC model, i.e. two
massless quarks in the color superconducting phase, whose main
features are as follows. At zero density the theory is invariant
under the group $\dd SU(3)_c\times SU(2)_L\times SU(2)_R$,  but at
high density Cooper pairing of two quarks  is energetically
favored \cite{alford}. The condensation from single-gluon exchange
between two quarks takes place in the color antitriplet channel.
The condensate breaks $\dd SU(3)_C$ down to an $\dd SU(2)_C$
subgroup and therefore  the quarks with nontrivial $\dd SU(2)_C$
charges acquire a gap $\dd \Delta$; moreover five gluons acquire
mass by the Higgs mechanism. The chiral symmetry remains unbroken,
which implies that there are no goldstone bosons.  We will use
here an effective theory near the Fermi surface \cite{hong}, which
has been recently applied to the CFL phase \cite{cfl} and to the
crystalline phase \cite{2fla} (for a discussion of the
superconductive crystalline phase see \cite{loff}). The main
results of this approach are summarized in Section 2. In Section 3
we compute the gluon self-energies; we confirm the results
obtained by other authors \cite{rischke1}, \cite{son} for the
masses of the five gluons
 associated to the
broken generators and for the dispersion laws of the unbroken
gluons; moreover we extend these results to the dispersion laws of
all the gluons. In Section 4 we present the one-loop corrections
to the three and four gluon vertices in the 2SC phase for all the
eight gluons, which allows for a complete effective description of
the gluon degrees of freedom of this phase at zero temperature.
%5, 6 and 7 and of the time component of the gluon 8 are of order
%$\Delta$, whereas the spatial components of the gluon 8 has a rest
%mass of order $g\mu$, due to the non renormalization of the time
%derivative of the corresponding field.
%We find that the dynamics of $\dd 4$ of these gluons is
%unaffected by the characteristics of the medium, whereas the
%longitudinal mode of gluon $\dd 8$ behaves similary to the
%massless gluons.

\section{Effective Theory}
To start with let us recall some results valid for the 2SC model.
In ref. \cite{hong}, \cite{cfl} an effective two dimensional field
theory for the CFL phase of QCD in terms of velocity dependent
fields was developed; in particular in  \cite{cfl} it was applied
to the computation of the gluon dispersion laws. In \cite{2fla}
this theory was extended to the 2SC model (in the crystalline
phase). The main ideas are as follows. To describe excitations
near the Fermi surface one writes the momentum $\dd p$ of the
quarks as \be p^{\nu} \ = \mu v^{\nu} + l^{\nu}~, \ee where $\dd
\mu$ is the quarks chemical potential and $\dd  v^{\nu} = (0,\vec
v_F)$, with $\dd \vec v_F$ Fermi velocity ($\dd |\vec v_F|=1$).
Only positive energy states $\psi_+$ contribute to the lagrangian,
whereas negative energy states decouple and can be expressed in
terms of the positive energy states. If we define \be
\gamma^\mu_\perp~=~\frac 1 2 \gamma_\nu\left(2 g^{\mu\nu}-
V^\mu\tilde V^\nu-\tilde V^\mu V^\nu \right)
 ~, \ee
\be V^\mu=(1,\vec v_F),~~~\tilde V^\mu=(1,-\vec v_F)~, \ee we can
write the negative energy states as
 \be
\psi_-=-\frac{1}{2\mu}\gamma_0\slash{\partial}_T\psi_+\ , \ee
showing the decoupling of $\dd \psi_-$ in the $\dd \mu\to\infty$
limit. Expressing $\dd \psi_-$ in terms of $\dd \psi_+$ results in
an effective theory which at the next to leading order in teh
inverse of $\dd \mu$ is described by the lagrangian \be
 {\cal L}=\sum_{\vec v_F}
\left[\psi_+^\dagger iV\cdot D\psi_+  -
\frac{1}{2\mu}\psi_+^\dagger( \slash{D}_\perp)^2\psi_+\right]~,
\label{5} \ee where $D_\mu$ is the covariant derivative with
respect to the color group. At this stage it is useful to use a
different basis for the fermion fields. We introduce the six
fields  $\varphi_{+}^A \ (A=0,\cdots,5)$ by the formulae \bea
\psi_{+,i\alpha }&=& \sum_{A=0}^3\frac{(\sigma_A)_{i\alpha}}{\sqrt
2}\varphi_{+}^A ~~~~~~~~(i,\,\alpha=1,\,2)~,\cr
\psi_{+,13}&=&\varphi_{+}^4 ~,\cr \psi_{+,23}&=&\varphi_{+}^5\
~,\eea where $\dd \sigma_A$ are the Pauli matrices for $\dd
A=1,2,3$ and $\dd \sigma_0=1$. The greek indices $\alpha,\beta$
are color indices and the latin indices $i,j$ are for the two
flavors $1,2$. Here clearly $\dd \varphi_{+}^A$ are positive
energy, velocity dependent fields: \be \varphi_{+}^A \equiv
\varphi_{+,\vec v}^A ~. \ee We also introduce the positive energy
fields with opposite velocity: \be \varphi_{-}^A \equiv
\varphi_{+,-\vec v}^A ~. \ee By defining \be \chi^A\
=\left(\begin{array}{c}
  \varphi_{+}^A  \\ {}\\
  C\varphi_{-}^{A\,*}\
\end{array}\right) ~,
\ee the lagrangian can be written as follows ($F^{a}_{\mu \nu}$
are the eight gluon fields): \be {\cal L}= \frac{1}{2} \sum_{\vec
v_F} \sum_{A=0}^{5} \chi_A^\dagger \left[ \matrix{iV \cdot D -
\dd\frac{1}{2\mu} \slash{D}^2_\perp  & \Delta^A\cr\Delta^A
&i\tilde V\cdot D^* - \dd\frac{1}{2\mu} \slash{D}^2_\perp} \right]
-\frac{1}{4} F^{\mu \nu}_{a} F_{\mu \nu}^{a} ~, \label{eff} \ee
which describes, at the lowest order, the effective theory. This
lagrangian allows for the evaluation of the two diagrams in Fig. 1
which give the one-loop contributions to the polarization tensor
$\dd \Pi^{\mu \nu}_{a b}(p)$ ($\dd a,b$ are color $\dd SU(3)$
indices, $a,b=1,\cdots,8$).
\begin{figure}[htb]
\vskip.6cm\epsfxsize=6truecm
\centerline{\epsffile{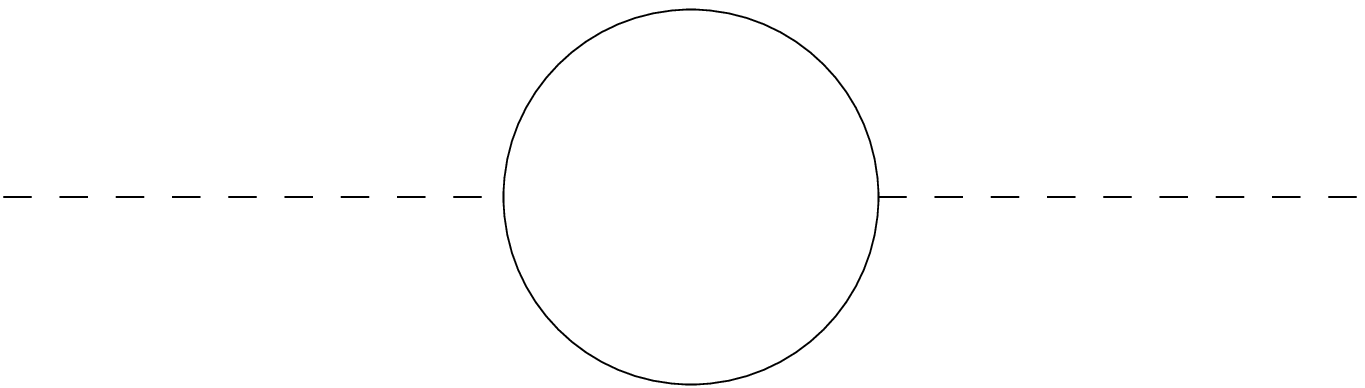}\hskip1cm\epsfxsize=6truecm\epsffile{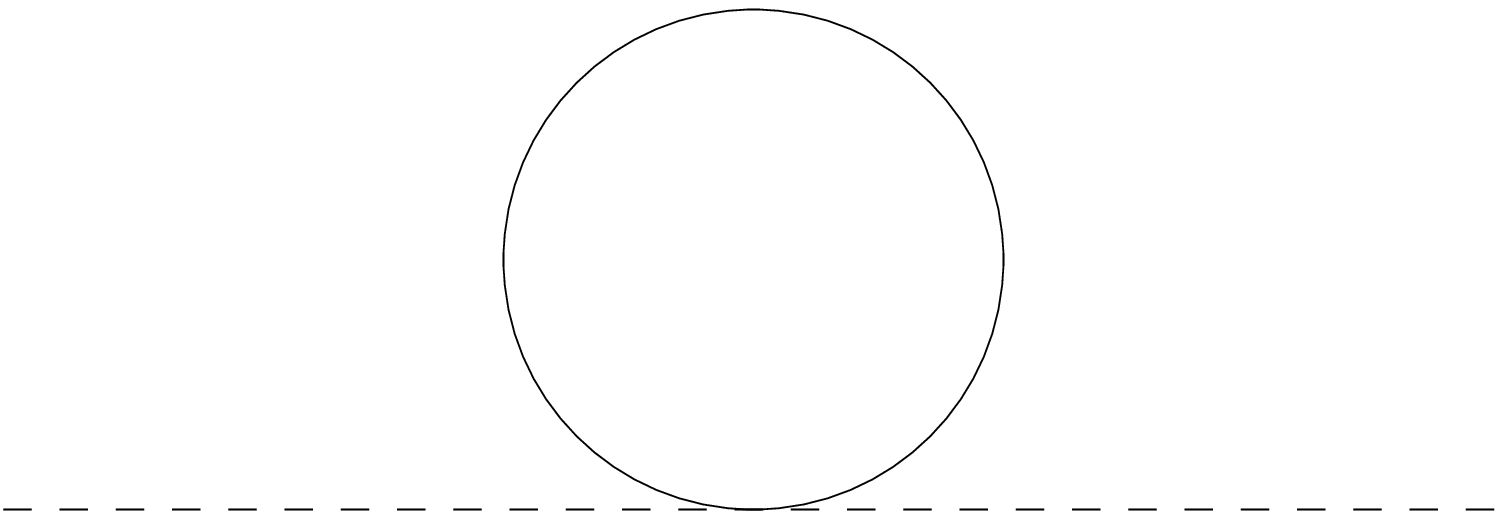}}
 \vskip1cm\noindent { Fig. 1}
{Gluon self energy diagrams; in both a) and b) dotted lines
represent gluon fields and full lines are  the fermion
propagators; b) is the tadpole diagram.}
\end{figure}
As discussed in \cite{cfl} the integration over the loop variable
$\ell$ is 2-dimensional ($\ell_0$, $\ell_{\|}$), while the
integration over directions perpendicular to the Fermi velocity
gives a factor of $\dd \mu^2/\pi$. The diagrams in Fig. 1 are the
only diagrams not suppressed in the $ \mu \rightarrow \infty$
limit. We write \be \Pi^{\mu \nu}_{a b}(p)\ =\ \Pi^{\mu \nu}_{a
b}(0) + \delta \Pi^{\mu \nu}_{a b}(p),  \ee with \be \Pi^{\mu
\nu}_{a b}(0)\ =\ \frac{\mu^2 g^2}{2 \pi^2} \int \! \frac{d\vec
v_F}{4 \pi} \Sigma^{0,\mu \nu}_{ab}, \label{pi1} \ee and \be
\delta \Pi^{\mu \nu}_{a b}(p)\ =\ \frac{\mu^2 g^2}{2 \pi^2} \int
\! \frac{d\vec v_F}{4 \pi} \Sigma^{\mu \nu}_{ab}(p). \ee To start
with we consider the diagram 1 a). Let us introduce the following
notations. We will denote the indices $a,b$ by the letters $i,j$
for $a,b=1,2,3 \,$, while for $a,b=4,5,6,7$ we  use the greek
letters $\alpha,~\beta$. The results of our calculation are, for
$a,b=i,j ~(=1,2,3)$: \bea \Pi^{\mu \nu}_{i j}&=& i \delta_{ij}
\frac{g^2 \mu^2}{4 \pi^3} \int \! \frac{d\vec v_F}{4 \pi}
 \int \! d^2 \ell\, \Big( \frac{ V^{\mu}V^{\nu} \tilde V\cdot
 \ell\,
 \tilde V \cdot(\ell+p)
 + \tilde V^{\mu} \tilde V^{\nu}  V\cdot \ell\, V
\cdot(\ell+p)}{D_1(\ell+p) D_1 (\ell)} \,+\cr&&\cr&+& \Delta^2
\frac{V^{\mu} \tilde V^{\nu} + V^{\nu} \tilde V^{\mu}}{D_1(\ell+p)
D_1 (\ell)} \,\Big) \ . \eea For $a,b=\alpha,\beta$ the
polarization tensor  is \bea \Pi^{\mu \nu}_{\alpha \beta}& = &i
\delta_{\alpha \beta} \frac{g^2 \mu^2}{8 \pi^3} \int \!
\frac{d\vec v_F}{4 \pi}
 \int \! d^2 \ell \left( V^{\mu}V^{\nu} \tilde V \cdot\ell\, \tilde V
\cdot(\ell+p)
 + \tilde V^{\mu} \tilde V^{\nu}  V \cdot\ell \,V\cdot (\ell+p)
\right)\times\cr&\times&
  \left( \frac{1}{D_1(\ell+p) D_2 (\ell)}
+ \frac{1}{D_2(\ell+p) D_1 (\ell) }\right) \ .
\label{glue4-7} \eea Finally for
the gluon 8 we have: \bea \Pi^{\mu \nu}_{8 8} &=& i \frac{g^2
\mu^2}{12 \pi^3} \int \! \frac{d\vec v_F}{4 \pi}
 \int \! d^2 \ell \Big[\left( V^{\mu}V^{\nu} \tilde V \cdot\ell\, \tilde
  V\cdot (\ell+p)
 + \tilde V^{\mu} \tilde V^{\nu}  V\cdot \ell\, V\cdot (\ell+p)
\right)\times\cr&\times& \left( \frac{1}{D_1(\ell+p) D_1 (\ell)} +
\frac{2}{D_2(\ell+p) D_2 (\ell) }\right) - \Delta^2 \frac{V^{\mu}
\tilde V^{\nu} + V^{\nu} \tilde V^{\mu}}{D_1(\ell+p) D_1 (\ell)} \
\Big] \ .
\label{glue8} \eea In the low momentum limit we can expand the
polarization tensor for $a,b=i,j ~(=1,2,3)$ in the following way:
 \be \Sigma^{0,\mu \nu}_{ij}\ =\
\delta_{ij} \left( \frac{ \tilde V^{\mu} \tilde V^{\nu} + V^{\mu}
V^{\nu}}{2} -\frac{ \tilde V^{\mu} V^{\nu} + \tilde V^{\nu}
V^{\mu}}{2} \right) ~, \ee and \be \Sigma^{\mu \nu}_{ij}(p)\ =\
\delta_{ij} \left( \frac{V^{\mu} V^{\nu} (\tilde V \cdot p)^2 +
 \tilde V^{\mu} \tilde V^{\nu} ( V \cdot p)^2}{12 \Delta^2}  -
 \frac{ V^{\mu} \tilde V^{\nu} +
 \tilde V^{\mu} V^{\nu} }{12 \Delta^2} ( V \cdot p\ \tilde V \cdot p)
 \right). \label{sig2}
\ee The validity of this approximation will be discussed below.
\newline It follows from eqs. (\ref{pi1}) - (\ref{sig2}) that \be
\Pi^{0 0}_{i j}(p)\ =\ \Pi^{0 0}_{i j}(0) + \delta \Pi^{0 0}_{i
j}(p) =\ \delta \Pi^{0 0}_{i j}(p) =\ \delta_{ij}\frac{ \mu^2
g^2}{18 \pi^2 \Delta^2} |\vec p\,|^2\, , \label{glue1} \ee \be
\Pi^{k l}_{i j}(p)\ =\ \Pi^{k l}_{i j}(0) + \delta \Pi^{k l}_{i
j}(p) =\ \delta_{ij} \delta^{kl} \frac{ \mu^2 g^2}{3 \pi^2} \left(
1 + \frac{p_0^2}{6 \Delta^2} \right)\, , \label{glue2} \ee and \be
\Pi^{0 k}_{ij}(p)\ =\ \delta \Pi^{0 k}_{i j}(p) =\ \delta_{ij}
\frac{ \mu^2 g^2}{18 \pi^2 \Delta^2}
 p^0 p^k \,.
\label{glue3} \ee These results agree with the outcomes of
\cite{rischke1} and \cite{son}.\\ For $\dd a,b=\alpha , \beta$
($\dd \alpha , \beta = 4,5,6,7$) we find: \be \Sigma^{0,\mu
\nu}_{\alpha \beta}\ =\ \delta_{\alpha \beta} \left( \frac{
V^{\mu} V^{\nu} +
 \tilde V^{\mu} \tilde V^{\nu}}{2} \right)
\ee and \be \Sigma^{\mu \nu}_{\alpha \beta}(p)\ =\ \delta_{\alpha
\beta} \left( V^{\mu} V^{\nu} \frac{(\tilde V \cdot p)^2}{4
\Delta^2} +
 \tilde V^{\mu} \tilde V^{\nu} \frac{( V \cdot p)^2}{4 \Delta^2} \right).
\ee After integrating over the Fermi velocities we obtain \be
\Pi^{0 0}_{\alpha \beta}(p)\ =\ \Pi^{0 0}_{\alpha \beta}(0) +
\delta \Pi^{0 0}_{\alpha \beta}(p)\ =\ \delta_{\alpha
\beta}\frac{\mu^2 g^2}{2 \pi^2} \left(1 + \frac{p_0^2+ |\vec
p\,|^2/3}{2 \Delta^2} \right), \label{pi004} \ee
 \be \Pi^{0 i}_{\alpha \beta}(p)\ =\ \delta \Pi^{0 i}_{\alpha \beta}(p)\ =\
 \delta_{\alpha \beta}
\frac{\mu^2 g^2}{6 \pi^2 \Delta^2} p^0 p^i ~, \ee and \be \Pi^{i
j}_{\alpha \beta}(p)\ =\ \Pi^{i j}_{\alpha \beta}(0) + \delta
\Pi^{i j}_{\alpha \beta}(p)\ =\ \delta_{\alpha \beta}\frac{\mu^2
g^2}{6 \pi^2} \left( \delta^{ij}+ \frac{\delta^{ij}p_0^2}{2
\Delta^2}+ \frac{ \delta^{ij} \vec p^2+2p^ip^j }{10 \Delta^2}
\right) . \label{piij4} \ee For $\dd a,b=8$ we get \be
\Sigma^{0,\mu \nu}_{88}\ =\ \left( \frac{5}{3} \frac{ V^{\mu}
V^{\nu} + \tilde V^{\mu} \tilde V^{\nu}}{2} + \frac{ V^{\mu}
\tilde V^{\nu} + \tilde V^{\mu}
 V^{\nu}}{6} \right) ~,
\ee \be \Sigma^{\mu \nu}_{88}(p)\ =\ \frac{1}{2} \left(
\frac{V^{\mu} V^{\nu} (\tilde V \cdot p)^2 +
 \tilde V^{\mu} \tilde V^{\nu} ( V \cdot p)^2}{9 \Delta^2}  +
 \frac{ V^{\mu} \tilde V^{\nu}
 +
 \tilde V^{\mu} V^{\nu} }{9 \Delta^2} ( V \cdot p\ \tilde V \cdot p)
 \right) \,;
\ee therefore we obtain \be \Pi^{0 0}_{88}(p)\ =\ \Pi^{0
0}_{88}(0) + \delta \Pi^{0 0}_{88}(p)\ =\ \frac{ \mu^2 g^2}{
\pi^2} \left(1 + \frac{p_0^2}{18 \Delta^2} \right), \ee and \be
\Pi^{0 i}_{88}(p)\ =\ \delta\Pi^{0 i}_{88}(p)\ =\ \frac{\mu^2
g^2}{54 \pi^2 \Delta^2} p^0 p^i \,, \ee \be \Pi^{i j}_{88}(p)\ =\
\Pi^{i j}_{88}(0) + \delta \Pi^{i j}_{88}(p)\ =\ \frac{\mu^2
g^2}{18 \pi^2} \left( 4 \delta^{ij}+ \frac{\delta^{ij} \vec
p^2+2p^ip^j}{15 \Delta^2} \right)\, . \ee These results complete
the analysis of Fig.1 a). Now we consider the diagram in Fig. 1
b).
 We note that this diagram is independent of the external
momentum $\dd p$, therefore it can only contribute to the gluon
masses. We also note that the diagrams of Fig. 1 present infrared
divergences (in $\ell_0$). To control these divergences one
considers the system in a heat bath and  substitutes the energy
euclidean integration
 $\dd \ell_4= - i \ell_0$ with a sum over the Matsubara frequencies
$\dd \ell_4 \rightarrow \omega_n = 2 \pi \left(n+\frac 1 2
\right)\beta$;
 eventually  one
performs the limit $T\ =\ \frac{1}{\beta}\ \rightarrow 0$. In this
way one finds for the contribution of the diagram 1 b) to $\dd
\Pi^{\mu \nu}_{a b}(0)$ the result ($a,b=1,\cdots,8$): \be
\Pi^{\mu \nu}_{a b}(0)\ =\ \frac{\mu^2 g^2}{4 \pi^2} \delta_{ab}
\int \frac{d\vec v_F}{4 \pi} \gamma^{\mu}_{T} \gamma^{\nu}_{T} ~.
\ee Therefore \be \Pi^{00}_{a b}(0)\ =\ 0 ~,\ee showing that there
is no contribution from this diagram to the Debye screening, while
one gets \be \Pi^{ij}_{a b}(0)\ =\ -\delta^{ij} \delta_{ab}
\frac{\mu^2 g^2}{3 \pi^2} ~. \ee In table 1 we summarize the
results for the Debye and Meissner masses obtained by the
calculations of the two diagrams in Fig.1. \vskip .2 true cm
\begin{center}
\begin{tabular}{|c|c|c|} \hline
  \hspace{.8in} & \hspace{.8in} & \hspace{.8in}  \\ [-.06in]
  $\dd a$     &  $\dd \Pi^{00}(0)$  & $\dd - \Pi^{ij}(0)$ \\ [.07in] \hline
                    &           &  \\ [-.09in]
$\dd 1 - 3$   &  $\dd  0$  &  $\dd 0 $ \\ [.07in] \hline
                   &           &  \\ [-.09in]
$\dd 4 - 7$ & $ \frac{3}{2} m^2_g $  &  $ \frac{1}{2} m^2_g$
\\[.07in] \hline
                   &           &  \\ [-.09in]
$\dd 8$   &  $ 3 m^2_g $& $ \frac{1}{3} m^2_g $\\[.07in] \hline
\end{tabular}
\vskip 0.3
 true cm {Table 1: Debye and Meissner masses for the gluons in the 2SC phase.}
\end{center}
\vskip 0.4 true cm \par\noindent where $\dd a$ is the gluon color
and $\dd m^2_g=\frac{\mu^2 g^2}{3 \pi^2}$ is the squared gluon
mass. Our results are in agreement with a calculation
performed by \cite{rischke1} with a different method.\\
\section{Dispersion law for the gluons}
In this section we will compute the dispersion laws for the
gluons. We begin our discussion by considering the unbroken colors
$a,b=i,j ~(=1,2,3)$.
\subsection{Gluons 1, 2, 3}
In this case we reobtain, by the present method, the results
already found in \cite{son} by a different approach, i.e. \be
{\mathcal L}\ =\ -\frac{1}{4} F^{\mu \nu}_{i} F_{\mu \nu}^{i} +
\frac{1}{2} \Pi^{\mu \nu}_{i j} A_{ \mu}^i A_{ \nu}^j\ ~, \ee with
$\Pi^{\mu \nu}_{i j}$ discussed above. Introducing the fields $\dd
E_i^a \equiv F_{0i}^a$ and $\dd B_i^a \equiv i \varepsilon_{ijk}
F_{jk}^a$, and using (\ref{glue1}), (\ref{glue2}) and
(\ref{glue3}) these results can be written as follows \be
{\mathcal L}\ =\  \frac{1}{2} (E_i^a E_i^a - B_i^a B_i^a) +
\frac{k}{2} E_i^a E_i^a ~, \label{33} \ee with \be k= \frac{g^2
\mu^2}{18 \pi^2 \Delta^2} ~. \ee As discussed in \cite{son} this
means that the medium has a very high {\it dielectric constant}
$\dd \epsilon = k +1$ and a {\it magnetic permeability} $\dd
\lambda = 1$. The gluon speed in this medium is now \be
v=\frac{1}{\sqrt{\epsilon \lambda}} \propto \frac{ \Delta}{g \mu}
\ee and in the high density limit it tends to zero. As shown in
\cite{son} the one loop lagrangian (\ref{33}) assumes the gauge
invariant expression \be  {\mathcal L}\ =\ - \frac{1}{4} F^{\mu
\nu}_j F^{j}_{\mu \nu} ~~~~ (j=1,2,3) ~, \label{36} \ee provided
the following rescaling is used \be A_0^j \rightarrow A_0^{j
\prime} =  k^{3/4} A_0^j ~, \label{A0} \ee \be A_i^j \rightarrow
A_i^{j \prime} =  k^{1/4} A_i^j ~, \label{Ai} \ee \be x_0
\rightarrow x_0^{\prime}  =  k^{-1/2} x_0 ~, \label{x0} \ee \be g
\rightarrow g^{\prime}  =  k^{-1/4} g ~. \label{g}
 \ee
 \subsection{Gluons 4-8}Let us now consider the equations of motion in
momentum space for the gluon field $\dd A_{\mu}^b$, $b=4,5,6,7,8
$: \be \left[\delta_{ab} \! \left(-g^{\mu\nu} p^2 + p^{\mu}
p^{\nu} \right) + \Pi^{\mu\nu}_{ab}\right]A_{\nu}^b\ =\ 0 ~.
\label{41} \ee We define the invariant quantities $\dd \Pi_0,
\Pi_1, \Pi_2$ and $\Pi_3$ by means of the following equations, \be
\Pi^{\mu \nu}(p_0,\,\vec p) = \left\{ \begin{array}{ll}
    \Pi^{00}=\Pi_0(p_0,\,\vec p)  \\
    \Pi^{0i}=\Pi^{i0}=\Pi_1(p_0,\,\vec p) \, n^i \\
    \Pi^{ij}=\Pi_2 (p_0,\,\vec p) \,\delta^{ij}\, +\, \Pi_3(p_0,\,\vec p)\,
n^i n^j
    \end{array}
    \right.
\label{dec} \ee where we have suppressed the color indices and
$\dd \vec n = \frac{\vec p}{p}$ . \\ For the broken degrees of
freedom it is useful to consider the scalar gluon field
$A_0^{a}$  and the longitudinal and transverse gluon fields
defined by \bea A_{i\, L}^{a} &=& \left( \vec n \cdot \vec A^a
\right) \, n_i ~, \cr A_{i\, T}^{a} &=& A_{i}^a-A_{i\, L}^{a}~.
\eea By the equation \be p_{\nu} \Pi^{\nu \mu}_{a b} A^b_{\mu} =
0 \ , \ee one obtains the relationship \be \left( p_0 \,\Pi_0\,
-\, |\vec p|\, \Pi_1 \right) A_0 = \vec n \cdot \vec A
 \left( p_0\, \Pi_1
\,-\,|\vec p|\, (\Pi_2+\Pi_3) \right) \ , \label{7} \ee between the scalar and
the longitudinal component of the gluon fields. The dispersion
laws for the scalar, longitudinal and transverse gluons are
respectively
\bea
\left(\Pi_2\,+\,\Pi_3
\,+\,p_0^2\right)
\left(|\vec p|^2\,+\,\Pi_0
\right)&=&p_0|\vec p|
\left(2\,\Pi_1\,+p_0|\vec p|\right)\ ,\cr
\left(\Pi_2\,+\,\Pi_3\,+\,p_0^2\right)
\left(|\vec p|\,p_0\,+\,\Pi_0\right)&=&p_0|\vec p|
\left(2\,\Pi_1\,+\,p_0^2\right)+\Pi_1^2
\ ,\cr
p_0^2-|\vec p|^2+\Pi_2&=&0\ .\label{pig}
\eea
Expanding
(\ref{7}) at first order in $\dd p$ we find for $\dd b= \beta =
4,5,6,7$ \be p^0 A_0^{\beta}\ =\ \frac{1}{3} \, \vec p \cdot \vec
A^{\, \beta} ~, \ee while for $\dd b=8$ \be p^0A_0^8\ =\
\frac{1}{9} \, \vec p \cdot \vec A^{\, 8}. \ee In any case we
define two masses, the rest mass: \be m^R\ =\ |p_0(|\vec
p|=0)| \ee  and the effective mass $\dd m^*$, by the formula: \be
\vec p\, = m^* \frac{ \partial E}{\partial \vec p}\, ~,\ee in the
low momentum limit. Using eqs. (\ref{pi004}) - (\ref{piij4}) we
observe that the effect of the kinetic terms in (\ref{41}) is
negligible at the order $\dd \frac{g^2 \mu^2}{ \Delta^2}$ for $\dd
\beta = 4,5,6,7$ in the high density limit. The dispersion law for
the time-like component is: \be \left( m_D^2 + \frac{m_D^2}{2
\Delta^2}(p_0^2+\frac{1}{3} |\vec p\,|^2) - \frac{m_D^2}{
\Delta^2} p_0^2 \right)A^{\beta}_0=0 ~, \ee where, from table 1,
we get $\dd m^2_D=\Pi^{00}(0)=\frac{\mu^2g^2}{2\pi^2}$; therefore
\be p_0= \pm E_0, \hspace{.2in} E_0= \frac{1}{\sqrt{3}} \sqrt{
|\vec p\,|^2 + 6 \Delta^2}. \ee The rest mass for these gluons, in
the gradient expansion approximation, is given by \be m^R_{A_{0}}\
=\ {\sqrt 2} \Delta . \label{m1} \ee This result shows that the
rest mass is of of the order of $\Delta$ and one could therefore
wonder if the result (\ref{m1}) is significant, since it is
obtained in the limit $|p/\Delta|\ll 1$.  To estimate the validity
of this approximation we use the exact result, which can be
obtained by Eqns. (\ref{pig}). Since $\Pi_3(p_0,0)=0$, the rest mass
of the three species
$A_0,\,A_L,\, A_T$ is given by \be m^2+\Pi_2(m^2,0)=0\
.\label{equa}\ee To obtain $\Pi_2(m^2,0)$ we integrate eq. (\ref{glue4-7})
, with $|\vec p|=0$, and we get
\be \Pi_2(m^2,0) =\ \frac{\mu^2g^2}{3\pi^2}
\left[-1+\int_0^{+\infty}  \! \! dx \, \frac{x +
{\sqrt{x^2+1}}}{(x+{\sqrt{x^2+1}})^2-(m/\Delta)^2} (1 -
\frac{x}{\sqrt{x^2+1} })
  \right]\ .
\ee
The numerical result of (\ref{equa}) is \be m\equiv m_R=0.894
\, \Delta \ . \ee A comparison with (\ref{m1}) shows that the
difference is of the order of $40-50\%$ and this is also the
estimated difference for the dispersion law  at small $\vec p$. We
notice that also in the three flavour case the gradient expansion
approximation tends to overestimate the correct result
\cite{shovkovy}. For the effective mass, we get in the gradient
expansion approximation\be \dd m^*_{A_{0}}\ =\ {\sqrt{18}}
 \Delta~. \label{m2} \ee
For the longitudinal and transverse cases we get respectively \bea
E_L^2+\frac{7}{15}|\vec p\,|^2&=&2 \Delta^2 ~, \\ E_T^2 +
\frac{1}{5}|\vec p\,|^2&=&2 \Delta^2 ~, \eea therefore the rest
masses are given by \be m^{R}_{A_L}=m^{R}_{A_T}=m^{R}_{A_0}=
{\sqrt 2} \Delta  ~.\ee The equality of the three rest masses is
an exact result, as we have stressed already.  On the other hand
the effective longitudinal and transverse masses are both negative
and their values are: \bea m^*_{A_L}&=&-\frac{15{\sqrt 2}}{7}
\Delta ~,\\ m^*_{A_T}&=&-5{\sqrt 2} \Delta ~. \eea We interpret
this result as follows: The spectrum of the quasi-particles
associated to these gluon modes has a maximum for $|\vec p| = 0$,
which means that at very small temperatures, which is the limit in
which we work, these quasi-particles are unlikely to be produced.
\newline For the time-like component of the gluon $\dd 8$ we have
the dispersion law \be E_{0}= \sqrt{9 \Delta^2  + 9  |\vec p\,|^2
\frac{ \Delta^2 }{ m_D^2} } ~, \ee where, in this case $\dd
m^2_D=\Pi^{00}(0)=\frac{\mu^2g^2}{\pi^2}$; therefore, \bea
m^R_{A_0}&=&3 \Delta ~,
\\ m^*_{A_0}&=&\frac{m^2_D}{ 3 \Delta}~.\eea For the longitudinal
and transverse modes we have respectively \be E_L\ =\ \sqrt{
\frac{4}{270}  \frac{m^2_M}{ \Delta^2}
 |\vec p\,|^2 + m_{M}^2} ~,\label{52}
\ee \be E_T\ =\ \sqrt{ - \frac{1}{30}  \frac{m^2_M}{ \Delta^2}
|\vec p\,|^2 + m_{M}^2} ~, \label{52b} \ee where the Meissner
 mass
is obtained from table 1: $\dd m^2_M=\frac{\mu^2g^2}{9\pi^2}$.
From these equations we see that \be m^R_{A_L}=m^R_{A_T}=m_{M} ~,
\label{54}\ee whereas the effective masses are as follows \bea
m^{*}_{A_L}&=& \frac{270}{4} \frac{\Delta^2}{m_M}~,\label{55}\\
m^{*}_{A_T}&=& - 30 \frac{\Delta^2}{m_M} ~.\label{55b} \eea We
note the peculiar feature of the spatial modes of the 8th gluon
that has a very large rest mass, see eq. (\ref{54}). This is an
exact result that is not obtained by the gradient expansion approximation.
Infact integrating eq. (\ref{glue8}), with $|\vec p|=0$, we get
\be
\Pi_2(m^2,\,0)\,=\,-\,\frac{\mu^2g^2}{9\pi^2}\,=\,-\,m_M^2\ , \ee
and therefore, from Eq. (\ref{equa}), one gets the result
(\ref{54}).

 The longitudinal and transverse gluons with
color 8 get nonetheless a vanishingly small effective mass, see
eqs. (\ref{55}) and (\ref{55b}), due to the very large coefficient
of $|\vec p|^2$ in (\ref{52}) and (\ref{52b}). These results
 should be contrasted with those obtained for the gluons
4, 5, 6, 7, and for the temporal mode of the gluon 8 (in all these
cases the rest mass is of order $\dd \Delta $). Since our
effective description is limited to energies  $< \Delta$  these
results mean that the longitudinal and transverse  modes of the
8th gluon are decoupled from the low-energy physics.
\section{Three and four point gluon vertices}
To complete the effective lagrangian for the eight gluons we
compute the one loop corrections to the gluon vertices $\dd
\Gamma_3$ (3 gluons) and $\dd \Gamma_4$ (4 gluons). We shall not
consider in the sequel a possible light glueball which is
discussed in \cite{sannino}. Therefore we write the full gluon
lagrangian as \be {\cal L}= -\frac{1}{4} F_{\mu\nu}^a F^{\mu\nu}_a
+ \frac{1}{2} \Pi^{\mu\nu}_{ab} A_{\mu}^a A_{\nu}^b + {\cal
L}^1_{(3)} + {\cal L}^1_{(4)} ~, \label{pippo} \ee where ${\cal
L}^1_{(3)}$ and  ${\cal L}^1_{(4)}$ are the one loop lagrangian
terms for the three and four point gluon vertices respectively.
For the three point gluon vertex we have two different diagrams
with a quark loop. The first one is depicted in Fig. 2; the second
one arises from the $1/\mu$ term in eq. (\ref{5}) and is
suppressed together with diagrams not containing quark loops. For
the four point gluon vertex we have three different diagrams with
a quark loop, but only the one depicted in Fig. 3 survives in the
$\mu\to\infty$ limit.
\begin{figure}[htb]
\epsfxsize=3.5truecm \centerline{\epsffile{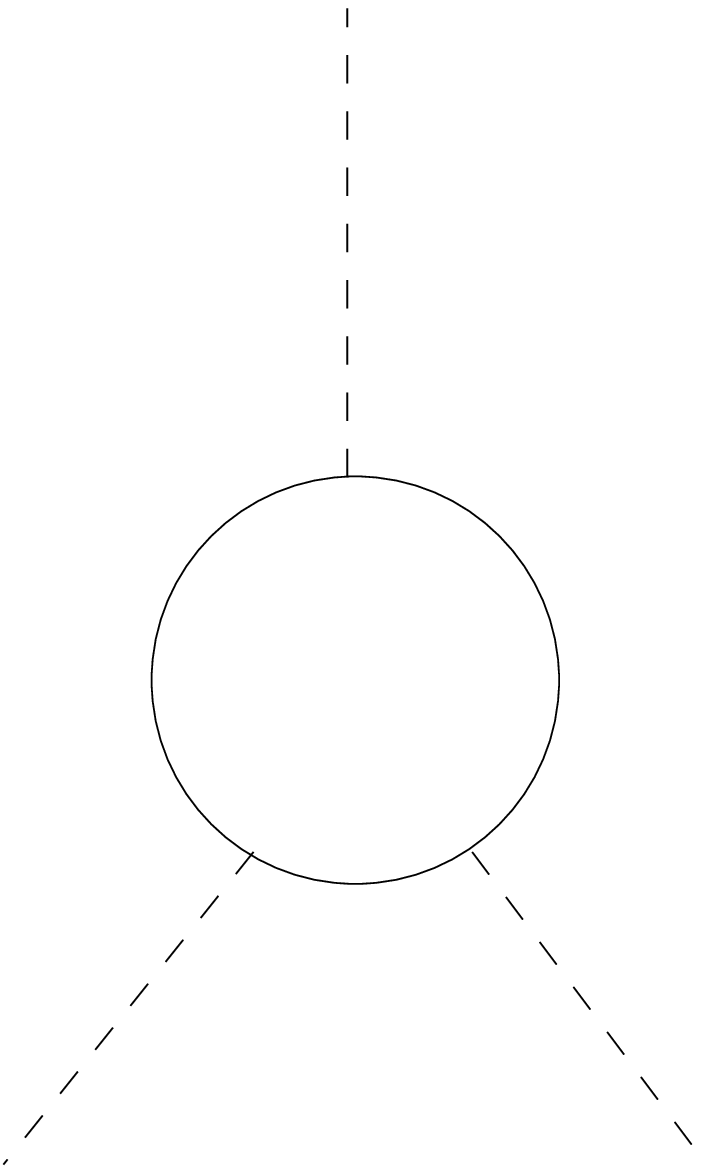}} \noindent
{Fig. 2}  {Three gluon vertex. Dotted lines represent gluon
fields; full lines are fermion propagators.}
\end{figure}
\begin{figure}[htb]
\epsfxsize=4truecm \centerline{\epsffile{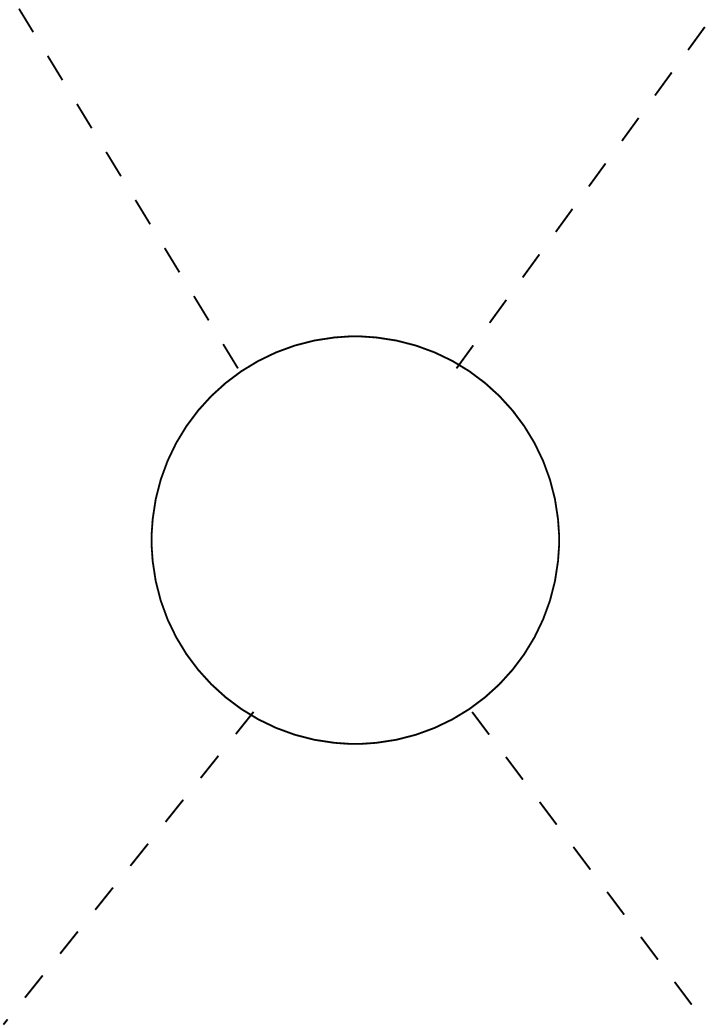}} \noindent {
Fig. 3}  {Four gluon vertex. Dotted lines  represent gluon fields;
full lines are fermion propagators.}
\end{figure}
\newline To start with, let us consider the three-point function. At the
tree level (${\cal L} = {\cal L}^0$) the contribution to the
lagrangian can be written in the form: \be {\cal L}_{(3)}^0 = -g
f_{abc} A_{a}^{\mu} A_{b}^{\nu}
\partial_{\mu} A_{c,\nu} ~. \label{L3} \ee At one loop
(${\cal L} = {\cal L}^1$) we have to distinguish between the
contribution of the diagram involving  the gluons in the unbroken
$SU(2)$ gauge group, which we call ${\cal L}_{(3),1}^1$, and those
involving  gluons corresponding to broken generators, which we
call ${\cal L}_{(3),2}^1$. So the one-loop correction at the three
gluon vertex may be written as follows: \be
 {\cal L}^1_{(3)}={\cal L}^1_{(3),1}+{\cal L}^1_{(3),2}.
 \ee
For the $\dd SU(2)$ contribution our results are as follows \be
{\cal L}_{(3),1}^1 = - g k f_{abc} A_{a}^{\mu} A_{b}^{\nu}
\de^{\mu} A_{c}^{\nu} [\delta_{\mu 0} \delta_{\nu i} + \delta_{\mu
i} \delta_{\nu 0}]~, \label{L31} \ee with $\dd a,b,c= i,j,l \in
\{1,2,3 \}$ and $\dd k=\frac{g^2 \mu^2}{18 \pi^2 \Delta^2}$. This
term can be obtained in a simpler manner, by requiring gauge
invariance for the $SU(2)$ gluons. On the other hand for $\dd
a,b,c = i,\alpha,\beta \, \ (i=1,2,3\, ;\ \alpha
,\beta=4,\cdots,7)$ or $a,b,c = 8,\alpha,\beta$ we have \be {\cal
L}_{(3),2}^1 = -g  \frac{3 k}{2} \left(I_{\mu \nu \rho \sigma}^1
C_1^{abc} + I_{\sigma \mu \nu \rho}^1 C_2^{abc} \right)
A_{a}^{\mu} A_{b}^{\nu}
\partial^{\sigma} \! \! A_{c}^{\rho} ~,
\label{L32} \ee where \be I_{\mu \nu \rho \sigma}^1 =\ \int \!
\frac{d\vec v_F}{4\pi} \, \tilde V_{ \mu} V_{ \nu} V_{ \rho} V_{
\sigma} \label{I1} \ee is the matrix which expresses the breaking
of Lorentz invariance. Its non-vanishing elements are listed in
table 2. \vskip .2 true cm
\begin{center}
\begin{tabular}{|c|c|c|c|c|c|}  \hline
  \hspace{.4in} & \hspace{.4in} &\hspace{.4in} & \hspace{.4in} &
\hspace{.4in} & \hspace{.4in} \\ [-.12in]
  $\mu \nu \rho \sigma$  & $0 0 0 0$ & $0 0 i i$ & $0 i 0 i$ & $0 i i 0$ &
$i i 0 0$ \\ [.12in] \hline
        &  &  & & &      \\ [-.12in]
  $I_{\mu \nu \rho \sigma}^1$  & $1$ &$ 1/3 $& $1/3$ & $1/3$ & $-1/3$ \\
[.12in] \hline \hline
        &  &  & & &      \\ [-.12in]
  $\mu \nu \rho \sigma$  & $i 0 i 0$ & $i 0 0 i$ & $i i j j$ & $i j j i$ &
$i j i j$ \\ [.12in] \hline
        &  &  & & &      \\ [-.12in]
  $I_{\mu \nu \rho \sigma}^1$  & $-1/3$ & $-1/3$ & $1/15$ & $1/15$ & $1/15$
\\ [.12in] \hline
\end{tabular}
\vskip 0.3
 true cm { Table 2: $I_{\mu \nu \rho \sigma}^1$ non-vanishing elements; $
i,j=1,2,~3$.}
\end{center}
\vskip 0.4 true cm \par\noindent On the other hand $C^{abc}_1$ and
$C^{abc}_2$ are the matrices which express the breaking of $\dd
SU(3)$ color and whose non-vanishing elements are given in the
table 3. \vskip .2 true cm
\begin{center}
\begin{tabular}{|c|c|c|} \hline
  \hspace{.6in} & \hspace{.6in} & \hspace{.6in} \\ [-.12in]
  $abc$   &  $C^{abc}_1$ & $C^{abc}_2$  \\ [.06in] \hline
                   &     &  \\ [-.12in]
 $\alpha 8 \beta $ &  $  \frac{1}{3} f_{\alpha 8 \beta} $ & $  \frac{2}{3}
f_{\alpha 8 \beta} $  \\ [.06in] \hline
                   &     & \\ [-.12in]
  $\dd \alpha i \beta$   & $f_{i \alpha \beta}$ & $0$ \\ [.06in] \hline
\end{tabular} \ ,
\vskip 0.3 true cm {Table 3: $C^{a b c}_1$ and $C^{a b c}_2$
non-vanishing elements.}
\end{center}
where $f_{abc}$ are the $SU(3)$ structure constants. Since the
coupling of the $\dd SU(2)$ gluons with gluons of color $\dd
\alpha=4,5,6,7$ is completely fixed by the gauge invariance, we
have written down the corresponding term $C^{i \alpha \beta}$ only
for a check. They are exactly the $SU(3)$ structure constants. \\
Let us consider the four-gluon vertex whose tree contribution is
\be {\cal L}_{(4)}^0 = - \frac{g^2}{4} f_{abe} f_{cde} A_{a}^{\mu}
A_{b}^{\nu} A_{c\,  \mu} A_{d\, \nu} ~. \label{L4} \ee The
one-loop correction may be written as follows: \be {\cal
L}_{(4)}^1 \ = {\cal L}_{(4),1}^1 + {\cal L}_{(4),2}^1 + {\cal
L}_{(4),3}^1 + {\cal L}_{(4),4}^1  . \ee We have four
contributions to ${\cal L}_{(4)}^1$. The first term is \be {\cal
L}_{(4),1}^1 = - k \frac{g^2}{4} f_{abe} f_{cde} A_{a}^{\mu}
A_{b}^{\nu} A_{c}^{\mu} A_{d}^{\nu} [\delta_{\mu 0} \delta_{\nu i}
+ \delta_{\mu i} \delta_{\nu 0}] ~, \label{L41} \ee with $\dd
k=\frac{g^2 \mu^2}{18 \pi^2 \Delta^2}$. It comes from diagrams
where all the gluons are in the $SU(2)$ group, therefore the
indices $\dd a,b,c,d$ take  the values $\dd i,j,l,m$. The second
term comes from diagrams with gluons of colors $\dd
a,b,c,d=\alpha,\beta,8,8$. The contribution to the lagrangian is
as follows: \be {\cal L}_{(4),2}^1 = -\frac{g^2}{4} \left(\frac{9
k}{2} \right) D^{abcd}_1 I^2_{\mu \nu \rho \sigma} A_{a}^{\mu}
A_{b}^{\nu} A_{c}^{\rho} A_{d}^{\sigma} ~. \label{L411} \ee  The
non-vanishing elements of the tensor $\dd D^{abcd}_1$ are in table
4 and \be I_{\mu \nu \rho \sigma}^2 =\ \int \! \frac{d\vec
v_F}{4\pi} \, V_{ \mu} V_{ \nu} V_{ \rho} V_{ \sigma} \ .\ee The
third contribution is from diagrams with gluons of colors  $\dd
a,b,c,d = i,\alpha,\beta,8$, with the gluon $8$ connected to two
ungapped quarks. The contribution is \be {\cal L}_{(4),3}^1 =
-\frac{g^2}{4} (9k) D^{abcd}_2 I^1_{\mu \nu \rho \sigma}
A_{a}^{\mu} A_{b}^{\nu} A_{c}^{\rho} A_{d}^{\sigma} ~,
\label{64}\ee where $I^1_{\mu \nu \rho \sigma}$ is the same tensor
we defined in equation (\ref{I1}) . The relevant $\dd D^{abcd}_2$
values are in table 4. The last contribution to the four-gluon
vertex is \be {\cal L}_{(4),4}^1 = -\frac{g^2}{4}
\left(\frac{9k}{2}\right) D^{abcd}_3 I^3_{\mu \nu \rho \sigma}
A_{a}^{\mu} A_{b}^{\nu} A_{c}^{\rho} A_{d}^{\sigma} ~,
\label{65}\ee where \be I_{\mu \nu \rho \sigma}^3 =\ \int \!
\frac{d\vec v_F}{4\pi} \, \tilde V_{ \mu} \tilde V_{ \nu} V_{
\rho} V_{ \sigma} \ .\ee and the non-vanishing $D^{abcd}_3$ are in
table 4.
 \vskip .2 true cm
\begin{center}
\begin{tabular}{|c|c|}  \hline
  \hspace{.6in} & \hspace{.6in} \\ [-.12in]
  $abcd$  &  $D^{abcd}_1$  \\ [.06in] \hline
        &           \\ [-.12in]
  $\alpha 8 8 \alpha$  &   $ \frac{4}{9} f_{\alpha 8 I}f_{8 \alpha I} $ \\
[.06in] \hline \hline
                   &     \\ [-.12in]
 $abcd$  &  $D^{abcd}_2$  \\ [.06in] \hline
        &           \\ [-.12in]
  $i \alpha 8 \beta$ & $\frac{2}{3} f_{ i \alpha I}f_{8 \beta I} $ \\
[.06in] \hline \hline
        &           \\ [-.12in]
  $abcd$  &  $D^{abcd}_3$  \\ [.06in] \hline
        &           \\ [-.12in]
  $i \alpha \beta j$ & $f_{ i \alpha I}f_{\beta j I} $ \\ [.06in] \hline
        &           \\ [-.12in]
  $8 \alpha \beta i$ & $\frac{1}{3} f_{ 8 \alpha I}f_{\beta i I} $ \\
[.06in] \hline
        &           \\ [-.12in]
  $\alpha \beta \gamma \delta$ & $f_{ \alpha \beta I}f_{\gamma \delta I}
  2 (1 + \log{4}) $ \\ [.06in] \hline
\end{tabular}
\vskip0.2cm{ Table 4: $D^{a b c d}$ non-vanishing elements.}
\end{center}
\vskip 0.2 true cm As we wrote above, some of these terms are
completely fixed by gauge invariance, once we know the
renormalization properties of the $SU(2)$ fields.\\ To evaluate
the effective lagrangian terms ${\cal L}_{(3)}$ and ${\cal
L}_{(4)}$ one might redefine the fields $A_{\mu}^a \ (a=1,2,3)$
and the coordinates according to eqs (\ref{A0}-\ref{g}); also the
other gluon fields can be redefined to include wave function
renormalization constants. We do not make this exercise since, for
the gluons corresponding to the broken colors, we gain little or
no physical insight from this procedure, as the corresponding
lagrangian cannot be put in the form (\ref{36}).

\section{Conclusions}
We have used the effective theory near the Fermi surface to
calculate the gluon self-energies and dispersion laws in the color
superconducting phase of QCD with two massless flavors (2SC). The
results confirm, within a different formalism, and extend, results
already obtained, notably by Rischke, Son, and Stephanov. The
three gluons of the unbroken $\dd SU(2)$ color have no Debye
screening and Meissner effect, but they are affected in their
dynamics by the medium polarizability. The remaining gluons show
Debye screening and Meissner effect. The Debye and Meissner masses
can be read from the table 1. For the gluons 1, 2, 3 one easily
finds the known result that they have no rest mass and no
effective mass and that their velocity is that relevant to a
polarizable medium of unit magnetic permeability and a dielectric
constant depending in a known way on the theory parameters. The
gluons 4, 5, 6 and 7 get masses of order of the gap, showing a
behaviour quite similar by the one exhibited by the gluons in the
CFL phase \cite{cfl}. The behaviour of the longitudinal gluon 8 is
not quite the same since there is no renormalization in the part
of the lagrangian involving the time derivative. As a consequence
the rest mass of this gluon is not of the order $\Delta$ (the
gap), but rather of order $g\mu$ (the Meissner mass). This makes
these particles to behave in-medium in a rather peculiar way. Very
difficult to be produced relatively to the other modes, because of
their large rest mass, but once produced they move as  particles
with a vanishingly small effective mass, of order $\dd
\Delta^2/g\mu$. The gluons 4, 5, 6, 7 have large (with respect to
the gap parameter) negative longitudinal and transverse effective
masses, so that they are unlikely to be produced at small
temperatures. To complete the construction of the effective
interaction we have calculated to one loop the three and four
point gluon interactions. Gauge invariance can be used directly
for the couplings involving gluons of the unbroken $\dd SU(2)$
color. For the remaining three and four point gluon selfcouplings
one has to evaluate explicitly the relevant loop diagrams. The
results are given in eqs (\ref{L3}-\ref{64}).

\end{document}